\documentclass[%
superscriptaddress,
preprint,
 amsmath,amssymb,
 aps,
]{revtex4-2}

\usepackage{graphicx}% Include figure files
\usepackage{dcolumn}% Align table columns on decimal point
\usepackage{bm}% bold math
\usepackage{color}
  
\usepackage[colorlinks, linkcolor=blue, anchorcolor=blue, citecolor=magenta]{hyperref}

\def \be {\mathbf{e}}

\def \bu {\mathbf{u}}

\def \bG {\mathbf{G}}

\def \bx {\mathbf{x}}
\def \bf {\mathbf{f}}
\def \br {\mathbf{r}}
\def \brs {\br_{\mathrm{s}}}
\def \brp {\br_{\mathrm{p}}}

\def \bq {\mathbf{q}}
\def \lp {\left(}
\def \rp {\right)}

\def \BF {B_{\mathrm{F}}}
\def \BL {B_{\mathrm{L}}}
\def \BR {B_{\mathrm{R}}}
\def \BS {B_{\mathrm{S}}}

\def \tbr  {\bar{\br}}
\def \tbrp {\tbr_{\mathrm{p}}}
\def \tbrs {\tbr_{\mathrm{s}}}
\def \tbus {\bar{\bu}_{\mathrm{s}}}

\def \bus {\bu_{\mathrm{s}}}

\def \tt {\bar{t}}
\def \tT {\bar{T}}
\def \td {\bar{d}_0}
\def \tE {\bar{E}}

\def \bes {\be_{\mathrm{s}}}

\def \tBF {\bar{B}_{\mathrm{F}}}
\def \tBL {\bar{B}_{\mathrm{L}}}
\def \tBS {\bar{B}_{\mathrm{S}}}
\def \tBR {\bar{B}_{\mathrm{R}}}
\def \d {\mathrm{d}}

\def \phip {\phi}
\def \brzero {\br_{0}}
\def \BF {B_{\mathrm{F}}}
\def \BL {B_{\mathrm{L}}}
\def \BS {B_{\mathrm{S}}}
\def \BR {B_{\mathrm{R}}}

\def \BFmax {B^{\mathrm{max}}_{\mathrm{F}}}
\def \BLmax {B^{\mathrm{max}}_{\mathrm{L}}}
\def \BSmax {B^{\mathrm{max}}_{\mathrm{S}}}
\def \BRmax {B^{\mathrm{max}}_{\mathrm{R}}}
\def \tUz {\tilde{U}^z}
\def \tLz {\tilde{L}^z}
\def \tUzoe {\tilde{U}_{\mathrm{eo}}^z}
\def \tLzoe {\tilde{L}_{\mathrm{eo}}^z}

\def \etaoe {\eta_{\mathrm{eo}}}
\def \etaot {\eta_{\mathrm{to}}}
\def \Toe {T_{\mathrm{eo}}}
\def \Tot {T_{\mathrm{to}}}
\def \Eoe {E_{\mathrm{eo}}}
\def \Eot {E_{\mathrm{to}}}
\def \Loe {L_{\mathrm{eo}}}
\def \Lot {L_{\mathrm{to}}}

\def \phip {\phi^{\prime}}

\def \thep {\theta^{\prime}}

\def \Lc {L_{\mathrm{clip}}}

\def \Ad {\hat{A}}

\begin{document}

\title{Optimising low--Reynolds--number predation via optimal control and reinforcement learning}% Force line 

\author{Guangpu Zhu}
\affiliation{Department of Mechanical Engineering, National University of Singapore, 117575, Singapore}

\author{Wen.-Zhen Fang}
\thanks{G. Zhu and W. -Z Fang contributed equally to this work.}%
\affiliation{Key Laboratory of Thermo-Fluid Science and Engineering, MOE, Xi’an Jiaotong University, Xi’an, 710049, China}
\affiliation{Department of Mechanical Engineering, National University of Singapore, 117575, Singapore}

\author{Lailai Zhu}
\email{lailai\_zhu@nus.edu.sg}
\affiliation{Department of Mechanical Engineering, National University of Singapore, 117575, Singapore}

\begin{abstract}
We seek the best stroke sequences of a finite--size swimming predator chasing a non--motile point or finite--size prey  at low Reynolds number. 
We use  optimal control  to seek the globally--optimal solutions for the former and  RL for general situations. The predator is represented by a squirmer model that can translate forward and laterally, rotate and generate a stresslet flow. We identify the predator's best squirming sequences to achieve the time--optimal (TO) and efficiency--optimal (EO) predation. For a point prey, the TO squirmer executing translational motions favours a two--fold L--shaped trajectory that enables it to exploit the disturbance flow for accelerated predation; using a stresslet mode significantly expedites the EO predation, allowing the predator to catch the prey faster yet with lower energy consumption and higher predatory efficiency; the predator can harness its stresslet disturbance flow to suck the prey towards itself; compared to a translating predator, its compeer combining translation and rotation is less time--efficient, and the latter occasionally achieves the TO predation via retreating in order to advance. We also adopt RL to reproduce the globally--optimal predatory strategy of chasing a point prey, qualitatively capturing the crucial two--fold attribute of TO path. Using a numerically emulated RL environment, we
explore the dependence of the optimal predatory path on  the size of prey. 
Our results might provide useful information that help design synthetic microswimmers such as \textit{in vivo} medical micro-robots capable of capturing and approaching objects in viscous flows. 
\end{abstract}

                              %display desired
\maketitle

\section{Introduction}~\label{sec:intro}
Approaching or chasing a moving target via optimal control has been a common task in natural and human settings, when e.g., animals like lions and sharks forage for prey, phagocytes chase and kill bacteria, predatory bacteria feed on other bacteria~\citep{perez2016bacterial,dashiff2011predation},  missiles intercept invading aircrafts, and shooters aim for running targets. The optimal foraging of natural creatures may still remain elusive, however, similar chasing applications in defense and robotic systems have become mature thanks to the optimal control theory.

Among these scenarios, the controlled agent and the target it approaches are in a dry environment such as air or in a liquid--filled wet environment. In the air, the agent and target, unless closely gaped or in a specific configuration (e.g., a missile in the wake of a high--speed aircraft), may not effectively affect the motion of each other by disturbing the air flow; namely, they can weakly sense the additional aerodynamic force induced by the motion of the other. In a liquid environment, the interaction between the agent and target is no longer weak because the viscosity of liquid is larger than that of air by three orders. Hence, a moving agent in liquid such as water can disturb its surrounding flow that exerts a considerable hydrodynamic force on its target nearby. This feature results in hydrodynamic interactions between the agent and target, which can significantly influence the chasing dynamics and the associated optimal chasing strategies. The effect of hydrodynamic interactions become especially pronounced and more long--ranged when the agent approaches/chases its targets in a low Reynolds number flow. This scenario commonly occurs for microorganisms or millimetre--scaled organisms swimming to approach motile particulate objects (\textit{e.g.}, bacteria or phytoplankton cells)
and non--motile counterparts such as organic debris~\citep{kiorboe2009mechanisms}. Both the type of swimmer and the size of target span a wide range~\citep{jabbarzadeh2018viscous}: a typical planktonic grazer is much larger than its prey~\citep{hansen1994size,kiorboe2016fluid}; an organism swims towards a similarly--sized member of the same species during bacterial conjugation~\citep{clark1962bacterial} or mating of copepods~\citep{strickler1998observing}; the target can also be much larger than the approacher exemplified by a spermatozoon swimming towards the egg or marine microbes targeting biological debris for  nutrients uptake and habitation~\citep{kiorboe2003marine}. Besides such natural events, similar situations might arise in the applications of future medical micro--robots, which need to approach targets such as bacteria and human cells of varying sizes~\citep{nelson2010microrobots,ceylan2019translational}. In these low--Reynolds--number flow configurations featuring important, long--ranged hydrodynamic interactions, prior explored predator--prey dynamics of territory/aerial animals or high--Reynolds--number aquatic animals together with the optimal chasing/approaching strategies would not apply. These tiny swimmers have evolved a variety of unique strategies suited for the viscous environment. For instance, zooplankton achieve feeding by means
of ambushing~\citep{kiorboe2009mechanisms}, generating currents~\citep{fenchel1980suspension}, cruising~\citep{kiorboe2009mechanisms} and colonizing marine snow aggregates.

A decent understanding of the predator--prey dynamics in viscous low--Reynolds--number flows would benefit analyzing the predatory and evasive behaviours of microorganisms, and exploring their evolutionary advantages. Besides, 
designing the optimal predatory and evasive strategies will be potentially useful in manipulating future medical micro--robots to capture bacteria or escape from hostile immune cells. 
Apart from substantial amount of studies on a related topic---nutrients uptake and feeding of swimming microorganisms~\citep{magar2003nutrient,langlois2009significance, tam2011optimal,michelin2011optimal,lambert2013active,kiorboe2014flow,dolger2017swimming,andersen2020effect},  work has addressed the interaction between a swimming predator and an individual particle or prey nearby. 
Without considering swimming--induced hydrodynamic effects, \citet{sengupta2011chemotactic} proposed and investigated a discrete chemotactic predator--prey model that describes a chasing predator and an escaping prey, which sense the diffused chemicals released from each other. \citet{pushkin2013fluid} studied theoretically and numerically the advection of a tracer and a material sheet of tracers when a microswimmer moves along an infinite, straight path. \citet{mathijssen2018universal} combined experiments, theory and simulations to perform a deep analysis of hydrodynamic entrainment of a particle by a swimming microorganism. Using a bispherical coordinate system, \citet{jabbarzadeh2018viscous} analytically studied the scenario of a forced spherical particle approaching another; they also numerically investigate the head--on approach of a self--propelling swimmer to another passive particle.
\citet{slomka2020encounter} conducted a modelling study on the ballistic encounter between elongated model bacteria and a  much larger marine snow particle that is sedimenting. Very recently, \citet{borra2022reinforcement} have studied a pair of point predator and prey considering their hydrodynamic interactions; they used a multi--agent reinforcement learning scheme to explore efficient, physically explainable predatory and evasive strategies.

In this work, we explore the optimal strategies of a finite--size swimming predator chasing a non--motile prey represented by a tracer point or a finite--size sphere.  The motion of the tracer is  purely driven by the propulsion-induced disturbance flow of the predator. Whereas for the spherical prey, we consider the two--way hydrodynamic coupling between the predator and prey based on numerical simulations. 
To seek the most time--saving or energy--efficient pursuing strategies of the predator, we adopt a numerical optimal control approach for a point prey and reinforcement learning (RL) for general cases. The RL--based optimal solutions qualitatively agree with and capture the essential features of the globally--optimal solutions identified by the former. We will demonstrate the emergence of non--intuitive optimal solutions in the seemingly simple configurations. We will also interpret physical mechanisms of the optimal strategies and discuss their implication on developing synthetic micro--robots designed for capturing moving objects.

\section{Problem setup, assumptions and methods}\label{sec:problem}
\subsection{Problem setup} 
\begin{figure}
\centering
\includegraphics[scale=0.8]{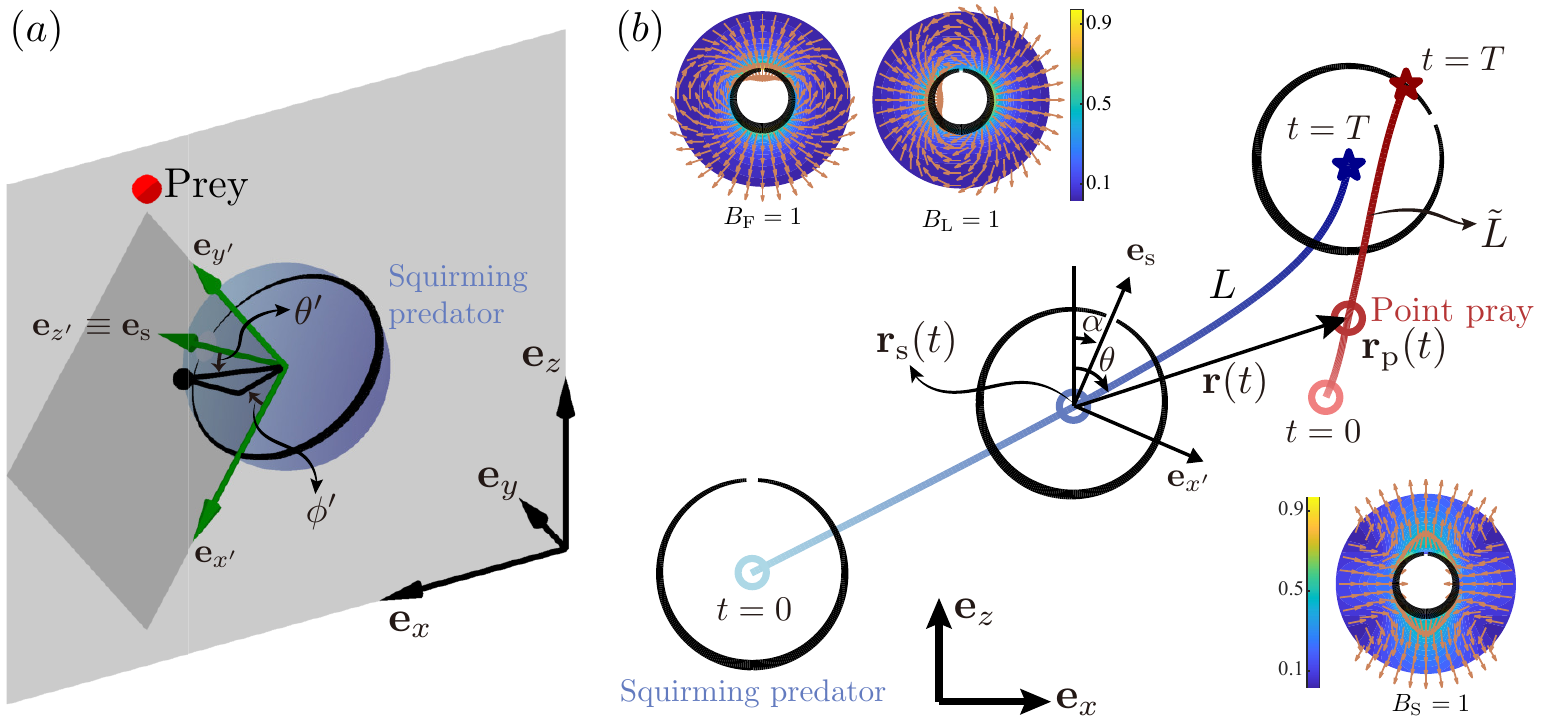}
\caption{(a), a squirming predator chases a point prey on the $y=0$ plane (finite--size spherical preys will be investigated in Sec.~\ref{sec:finite-size}).
$\be_{x^{\prime} y^{\prime} z^{\prime}}$ denotes the local coordinate system translating and rotating with the squirmer, while $\be_{xyz}$ is the global counterpart. (b), the predator captures the prey at time $t=T$, when the latter exactly touches the surface of the former. The coordinates of the squirmer center and prey are $\brs$ and $\brp$, respectively; $\br = \brp - \brs = r\lp \sin\theta\be_x + \cos\theta\be_z \rp$ indicates their relative displacement. The orientation $\bes$ of the squirmer deviates from $\be_z$ by an angle $\alpha$. The travelling distances of the predator and prey are denoted by $L$ and $\tilde{L}$, respectively. The insets show the disturbance velocity fields in the $y=0$ plane (in the lab frame) generated by a squirmer using the $\BF=1$, $\BL=1$ and $\BS=1$ modes.
The variables here are all dimensionless.}
\label{fig:sketch}
\end{figure}

We consider a microscale predator that swims to approach a prey in low Reynolds number flows (see Fig.~\ref{fig:sketch}(a)). A spherical squirmer of radius $A$ is adopted to model the predator, which
attains propulsion and rotation based on its surface actuation described by a slip velocity $\tbus(\thep,\phip)$, where $\thep \in [0,\pi]$ and $\phip \in[0,2\pi]$ are the polar and azimuthal angles with respect to the squirmer's swimming orientation $\bes$; here, $\bes$ coincides with $\be_{z^{\prime}}$ of the reference coordinates system $\be_{x^{\prime} y^{\prime} z^{\prime}}$ translating and rotating with the squirmer. The prey is modelled by a passively moving point tracer or a finite--size spherical particle of radius $\mathcal{A}$. The ratio $\chi =\mathcal{A}/A$ is defined to  indicate the size of prey compared to that of predator, which is zero for a point prey. From hereinafter, $\;\bar{}\;$ is used to denote dimensional variables. It is worth--noting that the squirmer model was proposed by \citet{lighthill1951,blake1971spherical}  for ciliary propulsion of microorganisms such as \textit{Paramecium} and \textit{Volvox}. This model has been successfully used to study microscale propulsion in the context of rheological complexity~\citep{Datt2015_JFM,lintuvuori2017hydrodynamics,li2021microswimming}, stratified fluids~\citep{more2020motion}, viscosity gradients~\citep{datt2019active}, effects of boundaries~\citep{spagnolie2012hydrodynamics,zhu2013low,ishimoto2013squirmer}, suspension of active particles~\citep{ishikawa2021rheology} and so on. 

Now we describe our predator--prey problem. At time $\tt = 0$, the orientation  $\bes$ of the squirming predator is in the $\be_z$ direction; the predator's center located at $\tbrs$ and the prey's center at position $\tbrp$ are in the $\bar{y}=0$  plane. To simplify the setting, we assume  a symmetric surface actuation velocity $\tbus$ about the plane, resulting in zero $\bar{y}$-component of the predator' swimming velocity and of its disturbance velocity in this plane. The latter implies that the prey will remain in the plane, as shown in Fig.~\ref{fig:sketch}(b); the angle $\alpha$ between the squirmer's orientation $\bes$ and the $\be_z$ axis indicates that $\bes = \sin{\alpha}\be_x + \cos\alpha \be_z$.
Besides, we assume that the predator detects the instantaneous position of the prey, following the previous works of intelligently controlled swimmers~\citep{gazzola2016learning,mirzakhanloo2020active}.
The predator is actuated by four squirming modes, $\tBF (\tt)$, $\tBL (\tt)$, $\tBR (\tt)$ and $\tBS (\tt)$; the first two modes allow it to translate forward and laterally with respect to its orientation $\bes$, respectively, corresponding to the $\be_{z^{\prime}}$ and $\be_{x^{\prime}}$ directions; $\tBR$ enables its rotation about the $\be_y$ axis; $\tBS$ introduces a stresslet flow. We would bound the strength of the surface actuations, namely, $\bar{B}_{i} \in [-\bar{B}^{\mathrm{max}}_{i},\bar{B}^{\mathrm{max}}_{i}], \; i=\mathrm{F},\mathrm{L},\mathrm{R},\mathrm{S}$. These modes will be described in their dimensionless form below.

 We choose  $A$ and $4 \bar{B}^{\mathrm{max}}_{\mathrm{F}}/A^3$ as the characteristic length and velocity, respectively.
To ease the calculation, we define the dimensionless displacement $\br=\brp - \brs$ between the predator and prey, which remains on the $y=0$ plane and can be described by its magnitude $r = |\br|$ and the angle $\theta$ between $\br$ and $\be_z$, namely, $\br = r\lp \sin\theta\be_x + \cos\theta\be_z \rp$.
The dimensionless slip velocity $\bus(\thep,\phip)$ on the surface of the squirmer in its own reference frame $\be_{x^{\prime} y^{\prime} z^{\prime}}$ reads~\citep{pak2014generalized}
\begin{align} \label{eq:slip_vel_non}
    \bus & =-\lp \frac{3\sin \thep}{2} \BF + \frac{3\cos \thep \cos \phip}{2} \BL + \frac{3\cos \phip}{4} \BR + \frac{9\sin 2\thep }{8} \BS \rp \be_{\thep} \\ \nonumber
    & + \lp \frac{3\sin \phip}{2} \BL + \frac{3 \cos \thep \sin \phip}{4} \BR\rp \be_{\phip}.
\end{align}

We will seek the best time sequences of the predator's surface actuations $[\tBF, \tBL, \tBR, \tBS]  (\tt)$ leading to the optimal predation. 
A standard optimum goal is to minimise the predating time $\tT$. We call such an optimization as the time--optimal (TO) optimization.
In addition to the capture time, we are also concerned about and will optimize the predatory energy efficiency $\eta$ defined as
\begin{align}\label{eq:eta}
    \eta = \frac{6\pi \mu  A \frac{\td}{\tT}  \times \td }{\tE},
\end{align}
where $\mu$ is the dynamic viscosity of the fluid, $\td = |\tbr(\tt=0)|-A-\mathcal{A}$ denotes the initial surface--to--surface distance between the prey and squirmer, and $\tT$ and $\tE$ represent the time and energy used by the predator to capture the prey, respectively. The numerator of Eq.~\ref{eq:eta} indicates the energy consumption of dragging a dead predator over a distance $\td$ to reach the prey within the capture time $\tT$. Using $4\pi \mu \bar{B}^{\mathrm{max}}_{\mathrm{F}}/3A$ as the characteristic energy scale, we write
\begin{align}
    \eta = \frac{6 d_0^2}{TE},
\end{align}
where $E=\int_0^T P(t) \d t$ and $P$ denotes the dimensionless power consumption of the squirmer. Accordingly, maximizing the energy efficiency $\eta$ is termed the efficiency--optimal (EO) optimization problem.

\subsection{Optimal control for a point prey}
We adopt a numerical optimal control approach when the prey is modelled by a point tracer, as described here.  The point prey is passively advected by the flow, hence, the translational and rotational velocities of the predator and the translational velocity of the prey in dimensionless form can be derived  as
\begin{subequations}\label{eq:vel_s_p}
\begin{align}
    \frac{\d \brs}{\d t} & =  \lp \BL \cos \alpha - \BF \sin \alpha \rp \be_x +  \lp -\BF \cos \alpha - \BL \sin \alpha \rp \be_z, \\
    \frac{\d \alpha}{\d t } & = \frac{3}{4} \BR, \\
    \frac{\d \brp}{\d t } & = \lp \frac{1}{r^3} \left[-\BF \cos (\theta - \alpha) + \BL \sin (\theta - \alpha) \right] + \frac{9}{8} \BS \frac{r^2-1}{r^4} \left[ 3 \cos^2(\theta - \alpha) - 1\right] \rp \be_r \\ \nonumber
    & + \lp \frac{1}{2 r^3} \left[ -\BL \cos (\theta - \alpha) - \BF \sin (\theta - \alpha) \right] - \frac{9}{8} \BS \frac{1}{r^4} \sin 2(\theta - \alpha) \rp \be_{\theta}.
\end{align}
\end{subequations}
Besides, the dimensionless power consumption of the squirmer reads
\begin{align}
    P = 12\lp \BF^2 + \BL^2 \rp + \frac{27}{2}\BS^2.
\end{align}
Since $d_0 = |\br(t=0)| - 1 -\chi = |\br(t=0)| - 1$ (realizing that $\chi=\mathcal{A}/A=0$ for a point prey) does not change over time, maximizing the predating efficiency $\eta$ is equivalent to minimizing the product of time $T$ and energy $E$. The optimal control problem for a point prey becomes: given an initial relative displacement $\br(t=0) = \brzero=r_0 \lp \sin\theta_0 \be_x + \cos\theta_0\be_z \rp $ with $|r_0|>1$ between the predator and the prey, the squirming predator will capture the prey at time $t=T$ when $|\br(t=T)|-1 \leq \varepsilon $, namely, the point prey is within a small cut--off distance $\varepsilon\ll 1$ from the predator's surface. The small parameter $\varepsilon$ is introduced here for theoretical convenience: when the prey moves very close to the squirmer's surface $|\br|\rightarrow 1$, the relative velocity between them will approach to zero because our squirmer only adopts the tangential but no radial surface actuation; hence, the prey will never exactly touch the squirmer's surface mathematically. In real situations, when they are sufficiently close, other physical ingredients would come into play; for example, diffusion via Brownian motion would allow them to touch each other~\citep{jabbarzadeh2018viscous}. 
In this work, we will use a fixed value of $\varepsilon=4\times10^{-3}$ unless otherwise specified. We have checked that varying $\varepsilon$ in the range $[10^{-3},10^{-1}]$ would
not alter the optimal chasing paths qualitatively; though the capture time will increase with decreasing $\varepsilon$ as anticipated.
Without loosing generality, the predator is initially oriented in the $\be_z$ direction, namely, $\alpha=0$ when $t=0$.
This predatory process corresponds to the evolution of $\br$ described by $r$ and $\theta$. Using $\frac{\d \br}{\d t} = \frac{\d \brp}{\d t}-\frac{\d \brs}{\d t}$ and Eq.~\ref{eq:vel_s_p}, we obtain the dynamical system characterizing the predatory process:
\begin{subequations}\label{eq:dynamical}
\begin{align}
    \frac{{{\d}r }}{{{\d}t}} =& \frac{{\lp { - 1 + {r^3}} \rp \cos \lp {\alpha  - \theta } \rp {\BF} + \lp { - 1 + {r ^3}} \rp \sin \lp {\alpha  - \theta } \rp{\BL}}}{{{r ^3}}} + \\ \nonumber
    &\frac{{9\lp { - 1 + {r ^2}} \rp\lp {1 + 3\cos \left[ {2\lp {\alpha  - \theta } \rp} \right]} \rp{\BS}}}{{16{r ^4}}}, \\
    \frac{{{\d}\theta }}{{{\d}t}} =& \frac{{\sin \lp {\alpha  - \theta } \rp\left[ {2r \lp {1 + 2{r ^3}} \rp{\BF} + 9\cos \lp {\alpha  - \theta } \rp{\BS}} \right] - \cos \lp {\alpha  - \theta } \rp\lp {2r  + 4{r ^4}} \rp{\BL}}}{{4{r ^5}}}.
\end{align}
\end{subequations}
We will seek the optimal sequences of the bounded actuation modes $[\BF,\BL,\BR,\BS](t)$ to minimize the capture time $t=T$ or to maximize the predating efficiency $\eta$ with these modes subject to:
\begin{subequations}
\begin{align}
    \BF  & \in [-1,1], \\
    B_i & \in [-B^{\mathrm{max}}_i,B^{\mathrm{max}}_i], \; i = \text{L}, \text{R}, \text{and } \text{S}. 
\end{align}
\end{subequations}
Unless otherwise specified, $\BLmax = \BRmax =\BSmax = 1$. This optimal control problem is solved numerically by an open--source library `FALCON.m'~\citep{falcon_web} implemented in MATLAB. The state variables are discretized in time by the trapezoidal collocation method. The nonlinear optimization problem is solved by the built--in open--source library IPOPT~\citep{wachter2006implementation}.

\subsection{Reinforcement learning for a point or finite--size prey}
Besides considering a passively moving point prey, we will also model the prey as a finite--size spherical particle 
of a dimensionless radius $\chi = \mathcal{A}/A$
that would hydrodynamically interact with the squirmer. The velocities of predator and prey cannot be derived analytically in the closed form as in Eq.~\ref{eq:vel_s_p}, and thus will be solved numerically. Accordingly, it is inconvenient to use the numerical optimal control approach as for the point prey, and we will instead adopt a deep RL scheme to identify the optimal predatory strategy. Naturally, the RL scheme can also be applied for the point prey model, as we will demonstrate in Sec.~\ref{sec:results_RL}.

We will extend an extensively validated  solver using boundary integral method (BIM) to emulate the hydrodynamic scenario of a swimming squirmer approaching a spherical prey. Different variants of the solver has been developed to study the micro--locomotion inside a tube~\citep{zhu2013low} or a droplet~\citep{reigh2017swimming}, dynamics of a particle--encapsulated droplet in shear flow~\citep{zhu2017bifurcation}, and a sedimenting sphere near a corrugated wall~\citep{kurzthaler2020particle}. A brief description of the BIM implementation in its dimensionless form is provided below.

In the spirit of BIM, we express the dimensionless velocity $\bu(\bx)$ at position $\bx$ everywhere in the domain as
\begin{equation}
\textbf{u}(\bx) = -\frac{1}{8 \pi} \int_{S_{\mathrm{s}}+S_{\mathrm{p}}} \bq(\textbf{x}')\cdot \bG (\textbf{x}, \textbf{x}')  \d S \lp \bx^{\prime} \rp,
\label{equation6}
\end{equation}
where $\bq$ is the the density of the so--called single--layer
potential on the surface $S_{\mathrm{s}}$ of the squirmer and that $S_{\mathrm{p}}$ of the prey. Here, $\bG(\bx,\bx^{\prime}) = \frac{\mathbf{I}}{|\bx-\bx^{\prime}|} + \frac{\lp \bx-\bx^{\prime}\rp\lp \bx-\bx^{\prime}\rp}{|\bx-\bx^{\prime}|^3}$ is the free--space Green’s function, which is also known as the Stokeslet.
Both the squirmer and finite--size prey are spherical, which can be discretized by zero--order quadrilateral elements. For either of the two, the hydrodynamic force and torque exerted on it are zero. This condition is used to determine their translational and rotational velocities.

Having introduced the BIM implementation, we now describe the RL algorithm. Compared to the optimal control theory that requires the predator--prey dynamics Eq.~\ref{eq:dynamical}, RL does not rely on prior knowledge of the dynamics but allows the squirmer as the predating agent to learn the dynamics, adapt and  optimize its chasing strategy (or policy in the language of RL) via continuously interacting with the environment. It is worth--noting that RL algorithms have been recently used
in similar swimming--involved scenarios, \textit{e.g.},
to optimize the swimming gaits or navigation routes of micro--swimmers at low Reynolds number~\citep{colabrese2017flow, schneider2019optimal,tsang2020self,mirzakhanloo2020active,muinos2021reinforcement,qiu2022navigation,nasiri2022reinforcement} and in turbulent flows~\citep{qiu2020swimming,alageshan2020machine}, macroscopic swimmers such as fish in viscous flows~\citep{gazzola2016learning,verma2018efficient} or in the potential flow~\citep{jiao2021learning}. Particularly, the recent pioneering experiments~\citep{muinos2021reinforcement} have demonstrated using RL for real--time navigation of micron-sized thermophoretic particles, opening a new horizon for developing swimming micro--robots endowed with artificial intelligence.

In this work, we adopt the open--source deep RL framework `Tensorforce'~\citep{tensorforce} and use a policy--based RL scheme---proximal policy optimization (PPO)~\citep{schulman2017proximal} to train the agent. 
The general idea behind the policy--based RL methods consists in parameterizing the policy function $\pi_{\Theta}$ by an artificial neural network (ANN) with a set of weights $\Theta$. The agent equipped with the parameterized policy identifies certain characteristic information, the state $s$, of the environment as the input to the ANN, and then selects an action $a$ according to the ANN's output. For the predator--prey system considered here, the state that the predator can observe is the position of the  prey relative to itself ($r$ and $\theta$), 
and the actions of the agent are the bounded squirming modes $B_i$. The selected action advances the environment from the current to the next state, and its effectiveness is 
quantified by an instantaneous reward $R$. 
An appropriate reward function shall favor the actions allowing the predator to approach the prey. 

To achieve the TO predation, we choose the following reward function:
\begin{align}
R = -r + \Gamma, \; \; \textrm{where} \;
   \Gamma = \left\{
   \begin{array}{lcr}
        -1000, &  &{r > 4}, \\
        200 / (t - \beta_{\text{T}}), &  &{r \leq (1+\varepsilon)},\\
        0,  &  &{\textrm{otherwise}},
   \end{array}
   \right.
\label{eq:R_r_Gamma}
\end{align}
where $t$ is effectively equivalent to the capture time $T$ and $\beta_{\text{T}}$ is the lower--bound estimation of $T$. Here, $-r$ promotes the predator to take only the necessary actions to 
approach the prey because any unnecessary ones will 
decrease the accumulated reward. The term $\Gamma$ contributes in two ways: first, it penalizes the agent for wandering far from the prey ($r > 4$) by activating a substantial negative reward; second, it
stimulates the predator to expedite the capture by offering a positive reward $\propto 1/\lp t-\beta_{\text{T}}\rp$ upon a successful capture. Note that we choose $\beta_{\text{T}} \approx d_0 $ as the initial surface--to--surface distance between the predator and prey.
For the EO setting, the  reward function is 
\begin{equation}
R = -r - \hat{\alpha} P + \Gamma, \; \; \textrm{where} \;
   \Gamma = \left\{
   \begin{array}{lr}
        -1000, & r > 4, \\
        20000 / \left[t E(t) - \beta_{\text{E}}\right], & r \leq (1+\varepsilon),\\
        0,  & \textrm{otherwise},
   \end{array}
   \right.
\label{eq:R_r_Gamma_En}
\end{equation}
where $\hat{\alpha}$ is a positive weight introduced to reduce the instantaneous power consumption $P$, and  $\hat{\alpha} = 0.1$ is chosen here. Also, $\beta_{\text{E}}$ is a lower--bound estimation of $ET$, chosen as half the value for a squirmer with $\BF=1$ travelling a distance of $d_0$.

Having defined the reward function, we describe the training process. The objective of RL here is to equip the agent with the optimal policy  maximizing the expected cumulative reward. After the agent executes the current ($k$--th) policy $\pi_{\Theta_k}$, we collect a set of trajectories $\{\nu_i\}$ (a trajectory is a sequence of states and actions, $\nu = \{s_0, a_0, ..., s_n, a_n, ...\}$) to determine the new [$(k+1)$--th] policy $\pi_{\Theta_{k+1}}$ via
\begin{equation}
\Theta_{k+1} = \textrm{arg} \; \underset{\Theta}{max} \underset{s, a \sim \pi_{\Theta_k}}{\mathbb{E}}[\Lc(s, a, \Theta_k, \Theta)].
\label{equation3}
\end{equation}  
Here, $\Lc$ is the so--called clipped surrogate advantage~\citep{SpinningUp2018} that measures the performance of a general policy $\pi_{\Theta}$ relative to the current one $\pi_{\Theta_k}$: 
\begin{equation}
\Lc[s, a, \Theta_k, \Theta] = min\left\{p(\Theta)\Ad^{\pi_{\Theta_k}}(s, a), \; \textrm{clip}\left[p(\Theta), 1-\hat{\epsilon}, 1+\hat{\epsilon} \right]\Ad^{\pi_{\Theta_k}}(s, a)\right\}.
\label{equation4}
\end{equation} 
Here, $\mathbb{E}$ denotes the mathematic expectation, $p(\Theta)= \pi_{\Theta}(a|s)/\pi_{\Theta_k}(a|s)$ is the probability ratio, and its numerator and denominator %$\pi_{\Theta_{k}}(a|s)$ and $\pi_{\Theta}(a|s)$
represent the probabilities of taking action $a$ in state $s$ at the $\pi_{\Theta}$ and  $\pi_{\Theta_{k}}$ policies, respectively; $\Ad^{\pi_{\Theta_k}}(s, a)$ is the advantage function, describing the advantage of choosing a specific action $a$ in state $s$ over that of a random choice  according to the  current policy $\pi_{\Theta_k}$. 
Besides, the function $\text{clip}\left[\right]$ indicates that $p \lp  \Theta \rp$ is  bounded in the range $\left[1-\hat{\epsilon},1+\hat{\epsilon}\right]$ via necessary clipping; here the hyperparameter $\hat{\epsilon}$ indicates how much the new policy is allowed to deviate from the current one and is fixed as $\hat{\epsilon}=0.2$ in this work.

\section{Results: Optimal control for a point prey ($\chi=0$)}\label{sec:results}
We start to investigate a point prey with $\chi=0$. In all cases, the squirming predator is initially aligned with the $\be_z$ axis, namely, $\alpha(t=0)=0$; the initial distance between the predator and prey is $|r_0|=3$, and $d_0=|r_0|-1=2$. We vary the initial bearing of the prey with respect to the predator, namely the angle $\theta-\alpha$ at $t=0$ between the predator--prey displacement vector $\br$ and the squirmer's orientation $\bes$ (see Fig.~\ref{fig:sketch}). This initial angle recovers to $\theta_0$ by realizing $\alpha(t=0)=0$. Here, $\theta_0=0^{\circ}$, $90^{\circ}$ and $180^{\circ}$ correspond to when the prey is in front of, on the right side and in the rear of the predator, respectively. 

\subsection{A predator combining forward and lateral motions}\label{sec:point_FL}
\begin{figure}
\centering
\includegraphics[scale=0.33]{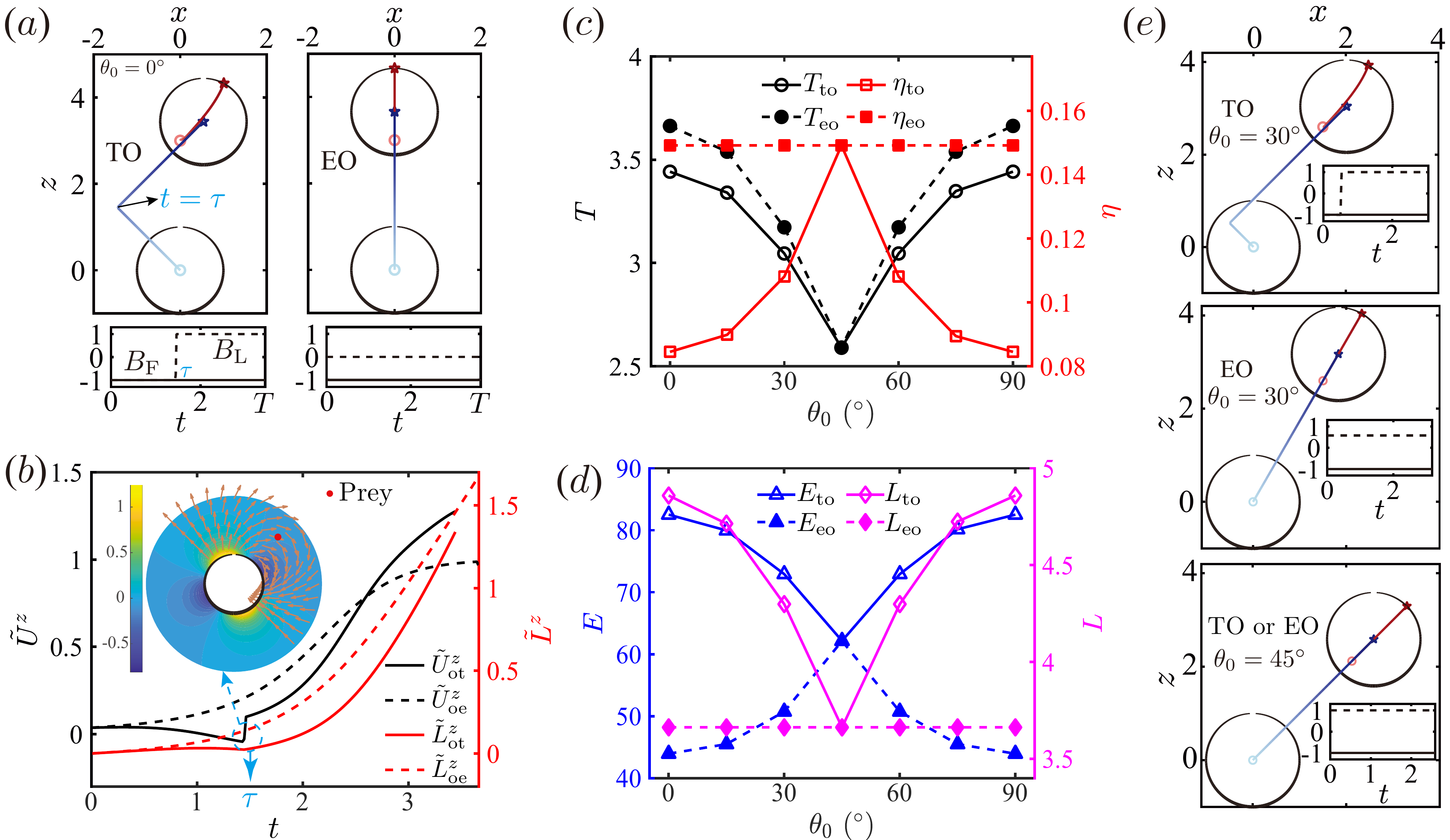}
\caption{Time--optimal (TO) and efficiency--optimal (EO) predating strategies using a combined forward and lateral squirming modes of magnitudes $\BF$ and $\BL$, respectively. (a), trajectories of the predator (blue) and point prey (red), where the latter is initially ahead of the former ($\theta_0=0^{\circ}$); circles and stars denote their initial and ending positions. The bottom panels show how the two modes evolve in time. The left and right panels depict the TO and EO strategies, respectively. (b), the prey' velocity $\tUz$ and accumulative travelling distance $\tLz$ in $\be_z$; $\tau$ indicates when the TO predator takes the sharp turn. The inset illustrates its disturbance flow field in the $y=0$ plane right before $t=\tau$. (c), capture time $T$ and predating efficiency $\eta$ of the TO and EO predators versus the initial bearing $\theta_0$ of the prey with respect to the predator. (d), similar to (c), but for predators' consumed energy $E$ and travelling distance $L$. (e), the TO and EO predating behaviours when $\theta_0 = 30^{\circ}$ and $45^{\circ}$.
}
\label{fig:OT_OE}
\end{figure} 

We first consider a predator swimming forward and laterally,  respectively, via the $\BF$ and $\BL$ squirming modes. This combination is termed as `F+L'. Without combining them, the predator with either of them alone can only swim vertically (in $\be_z$) or horizontally (in $\be_x$) and thus cannot reach the prey located in an arbitrary bearing $\theta_0$. Fig.~\ref{fig:OT_OE}(a) compares the predator and prey's trajectories of the TO and EO cases. The EO predator swims directly towards the prey by attaining the maximum forward movement ($\BF=-1$) and zero lateral movement ($\BL=0$), which follows an intuitive predatory strategy by taking the straight thus shortest path. In contrast, the TO predator chooses an L--shaped route oriented first in the northwestern direction then followed by a sharp $90$--degree turn towards the northeastern direction. Correspondingly, $\BF=-1$ producing the maximum forward motion holds during the whole course while $\BL$ jumps sharply from $-1$ to $1$ at the turning time $t=\tau$. 
We now discuss the mechanism behind this peculiar strategy. The predator initially lags behind the prey by a distance of $3$ in $\be_z$. Both the TO and EO predators adopt $\BF=-1$ to maintain the same maximum movement in that direction, the difference in the capture time $T$ then depends on the prey's velocity $\tUz$ in $\be_z$. Fig.~\ref{fig:OT_OE}(b) compares $\tUz$ of the two prey and their accumulative travelling distances $\tLz$ along the $\be_z$ direction. An important observation is that the EO prey's $\tUz$ remains positive indicating its consistent motion away from the predator, whereas the TO prey's velocity $\tUz$ becomes negative when $t<\tau$ implying its motion towards the predator. The inset of Fig.~\ref{fig:OT_OE}(b) depicts the instantaneous flow field around the squirmer right before $t=\tau$, reflecting the negative velocity $\tUz$ experienced by the prey (red dot). To sum up, the EO predator swimming straight towards the prey generates a flow field that always repels the prey away from it; while the TO predator adjusts its position (with respect to the prey) and surface actuation for best exploiting its disturbance flow field to attract the prey, leading to the initially left--upward movement. In addition to the $\theta=0^{\circ}$ orientation, similar trajectories of the OT and OE predators are found for an arbitrary orientation $\theta_0 \neq 45^{\circ}$ of the prey, as shown in Fig.~\ref{fig:OT_OE}(e).

We then examine how the initial orientation $\theta_0\in[0,90]^{\circ}$ of the prey with respect to the predator affects the predating dynamics; the results for $\theta_0\in[90,180]^{\circ}$ are not shown by realizing the fore--aft symmetry. We depict the capture time $T$ and efficiency $\eta$ of the TO and EO strategies in Fig.~\ref{fig:OT_OE}(c), and the predators' energy consumption $E$ and travelling distance $L$ in Fig.~\ref{fig:OT_OE}(d). All the quantities exhibit a mirror symmetry about $\theta_0=45^{\circ}$ when the prey is right in the northeastern direction. 
In this particular configuration, the TO and EO predators adopt the same strategy---swimming straight towards the prey, as shown by the bottom panel of Fig.~\ref{fig:OT_OE}(e). This symmetry can be anticipated because the forward $\BF$ or lateral $\BL$ mode alone allows the predator to approach the prey at $\theta_0 = 0^{\circ}$ or $90^{\circ}$, respectively. For the two modes sharing the same magnitude $\BFmax=\BLmax=1$, the predatory scenario for the $90-\theta_0$ orientation can be obtained by interchanging the time sequences of $\BF$ and $\BL$ for the $\theta_0$ counterpart. In addition, we see that the EO predators achieve an optimal efficiency of $\etaoe\approx0.15$ independent of $\theta_0$, which almost doubles that of the TO predator when $\theta_0=0^{\circ}$ and $90^{\circ}$. In contrast, the TO predator captures the prey slightly faster than the EO counterpart, reducing the capture time at most by around $0.2$, which occurs when the prey is ahead of or beside the predator. Both the TO and EO predators catch their prey fastest that are initially in the northeastern direction ($\theta_0=45^{\circ}$) and take more time when they deviate from that orientation. Besides, compared to the EO predator, the TO predator consumes more energy.

\begin{figure}
\centering
\includegraphics[scale=0.4]{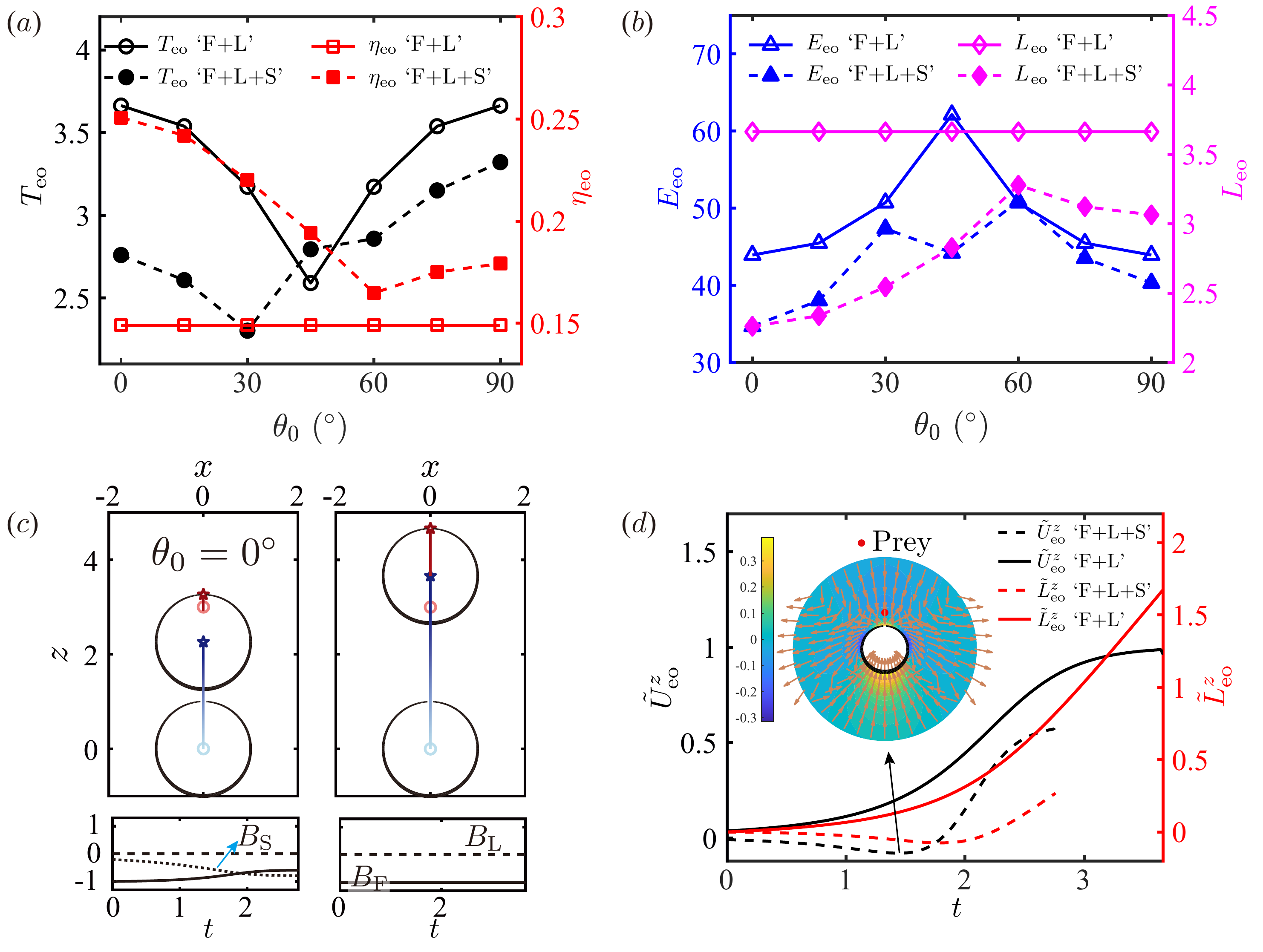}
\caption{Adding a stresslet mode $\BS$ on top of the EO predator with $\BF$ and $\BL$ modes 
enhances its all-round performance. (a): capture time $\Toe$ and predatory efficiency $\etaoe$ versus the initial bearing $\theta_0$ of the prey. (b): similar to (a), however, for the predators' energy consumption $\Eoe$ and travelling distance $\Loe$. (c), trajectories of the EO predator and prey with (left panel) and without (right panel) the $\BS$ mode; the bottom panels show the time sequences of the squirming modes. (d), the prey's velocity $\tUzoe$ and accumulative travelling distance $\tLzoe$ in $\be_z$, with and without the $\BS$ mode. The inset indicates the flow field when the $\tUzoe$ reaches the minimum.
}

\label{fig:BS}
\end{figure} 
\subsection{The stresslet squirming mode facilitates predation}
Having observed how the predator exploits its squirming--induced disturbance flow to catch the prey faster, we then examine the influence of the stresslet mode $\BS$ known to vary the disturbance flow without affecting the swimming speed of an isolated squirmer~\citep{blake1971spherical}. According to our definition (with a sign difference compared to classical definitions), $\BS<0$ corresponds to a puller micro--swimmer, e.g., the biflagellated
algae \textit{Chlamydomonas}; $\BS>0$ indicates a pusher counterpart exemplified by most flagellated micro--organisms. The puller attracts the fluid from its front and rear towards itself, while the pusher drives the flow oppositely. Compared to the baseline `F+L' predator using only the $\BF$ and $\BL$ modes, we show in Fig.~\ref{fig:BS} that introducing the stresslet mode $\BS$ can significantly enhance the all-round predatory performance under the EO policy. Fig.~\ref{fig:BS}(a) shows that the `F+L+S' (with $\BF,\BL$ and $\BS$ modes) predator captures the prey faster than the `F+L' (with $\BF$ and $\BL$ modes) competitor for all the bearings of the prey $\theta_0$, except for when $\theta_0=45^{\circ}$. Moreover, incorporating the stresslet considerably enhances the predatory efficiency $\etaoe$ of the EO predators with a maximum relative enhancement approaching $70\%$ when $\theta_0=0^{\circ}$, as well as decreases its energy consumption $\Eoe$ and travelling distances $\Loe$ for all the prey's initial bearings (see Fig.~\ref{fig:BS}(b)).  

In particular, we examine in Fig.~\ref{fig:BS}(c) the straight trajectories of EO predators with (left panel) and without (right panel) the stresslet mode together with those of their respective prey. We observe that the prey has moved from the stresslet--equipped predator by a negligible distance compared to that chased by the stresslet-free one. The former predator with a negative $\BS$ mode (shown in the left-bottom panel of Fig.~\ref{fig:BS}(c)) behaves as a puller swimmer, which significantly sucks the prey ahead towards it. The associated sucking disturbance flow generated by the predator is depicted in the inset of Fig.~\ref{fig:BS}(d). This mechanism of stresslet--accelerated predatory process is also confirmed by the evident backward motion---negative $\tUzoe$ and $\tLzoe$ of the prey shown in Fig.~\ref{fig:BS}(d). In retrospect, it was analogously found that biflagellated organisms can enhance their feeding performance by adopting  a puller--styled locomotory gait~\citep{dolger2017swimming}. As expected, when the prey is initially on the right side of the predator ($\theta_0=90^{\circ}$), the latter would activate a positive $\BS$ mode; accordingly, the prey is attracted laterally towards to pusher--styled swimmer, which is not shown here.

\begin{figure}
\centering
\includegraphics[scale=0.4]{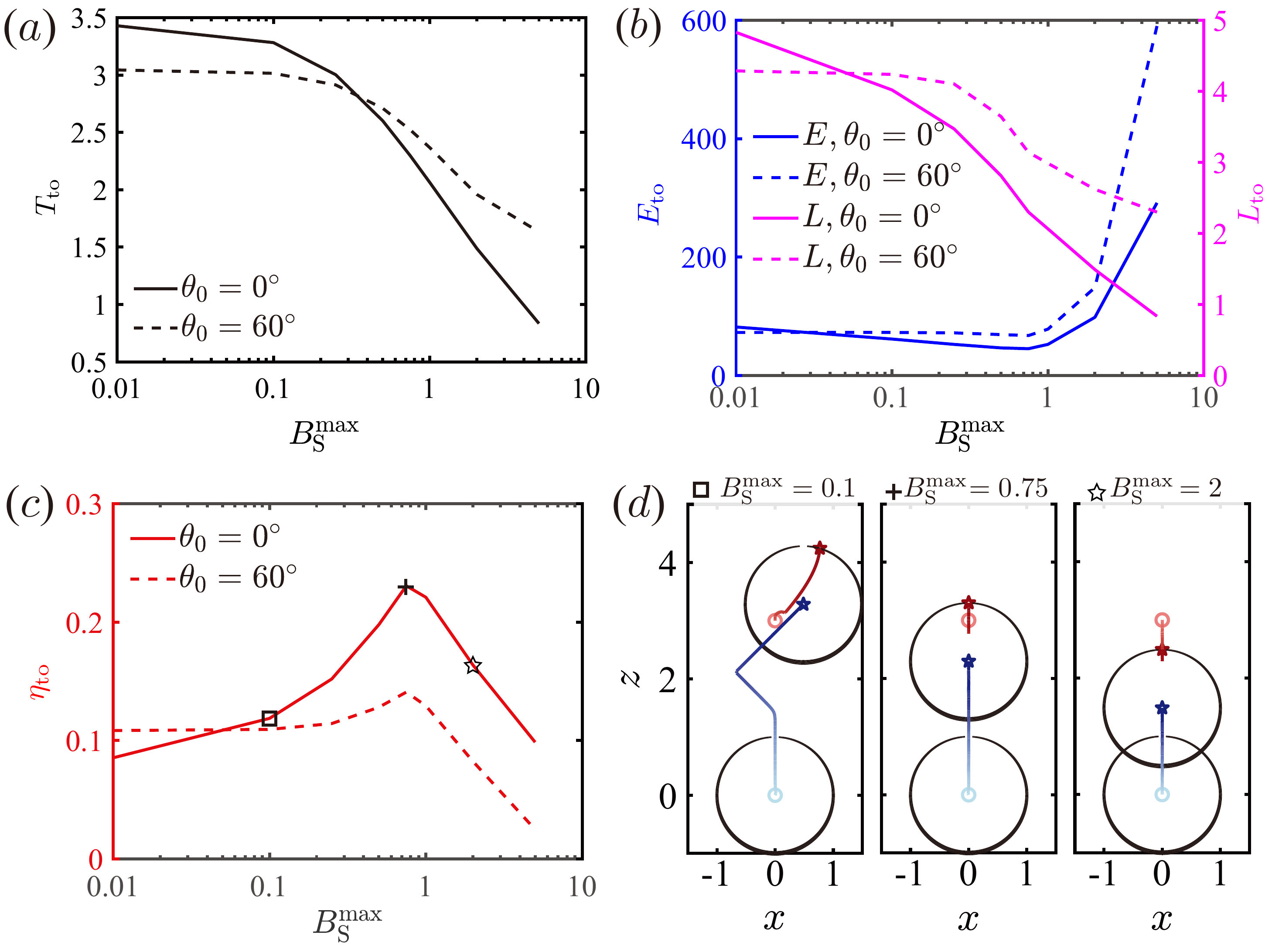}
\caption{Varying the maximum magnitude $\BSmax$ of the stresslet mode for the TO predator that swims using the `F+L+S' combination of squirming modes. (a) Capture time $\Tot$ versus $\BSmax$ for two initial orientations, $\theta_0=0^{\circ}$ and $\theta_0=60^{\circ}$, of the prey with respect to the predator. (b) Energy consumption $\Eot$ and travelling distance $\Lot$ versus $\BSmax$. (c) Similar to (a), but for the predating efficiency $\etaot$. (d) Trajectories of the predator and prey for  $\BSmax=0.1$, $0.75$ and $2$ marked in (c). }
\label{fig:varyBS}
\end{figure} 

Intuitively, we infer that a puller predator can exploit such a stresslet disturbance flow to accelerate capturing the prey. Hence, the time optimal predator would naively turn on the full gear of $\BS$ mode for the fastest capture.
This intuition is confirmed
by Fig.~\ref{fig:varyBS}(a) showing that $\Tot$ decreases monotonically with  increasing $\BSmax$ when $\theta_0 = 0^{\circ}$ and $60^{\circ}$. On the other hand, the growing stresslet will produce higher power consumption of the predation due to the stronger viscous dissipation of the fluid. Fig.~\ref{fig:varyBS}(b) depicts that the energy consumption $\Eot$  weakly decreases with $\BSmax$ when $\BSmax <1$ but sharply increases with  $\BSmax >1$. The slightly negative relation between $\Eot$ and $\BSmax<1$ are due to: first, in this regime, the major power consumption is not from the stresslet flow but from the forward and lateral motions of squirmer; second, the decreasing time $\Tot$ in this regime with $\BSmax$ tends to lower down the total energy. When $\BSmax$ keeps growing from around $1$, the stresslet--induced power becomes increasingly dominant, because the swimming power scales with these modes quadratically while the forward ($\BF$) and lateral ($\BL$) modes are bounded to $1$. 

Next, we show in Fig.~\ref{fig:varyBS}(c) that the predating efficiency $\etaot$ non-monotonically depends on $\BSmax$, attaining the maximum when $\BSmax$ is approximately $0.7 - 0.8$. This non-monotonic dependence can be explained by the variation of $\Tot$ and $\Eot$ according to $\BSmax$. We then examine in Fig.~\ref{fig:varyBS}(d) three characteristic chasing scenarios for the $\theta_0=0^{\circ}$ situation. In the case  $\BSmax=0.1$, the predator first approaches the prey straightforward. Then, it takes a two--fold zigzag path as a reminiscent of the TO predatory strategy in the absence of $\BS$ mode shown in Fig.~\ref{fig:OT_OE}(a). The initial straight chasing reflects the predator's tendency to utilize the $\BS$--induced flow for sucking the prey. Increasing $\BSmax$ to $0.75$ results in the optimal efficiency $\etaot \approx 0.23$ that exceeds double the efficiency of the $\BSmax=0.1$ and $5$ predators. The suction flow of this $\BSmax$ level is strong enough to overcome the forward flow generated by the $\BF$ mode. Hence, the prey initially moves backward towards and then moves forward as their distance is decreased. When $\BSmax$ grows to $2$, the stresslet--induced suction completely dominates over the forward flow, hence enabling the prey to continuously move backward till being captured.

\subsection{Incorporating rotation or only using translations?}
\begin{figure}
\centering
\includegraphics[scale=0.37]{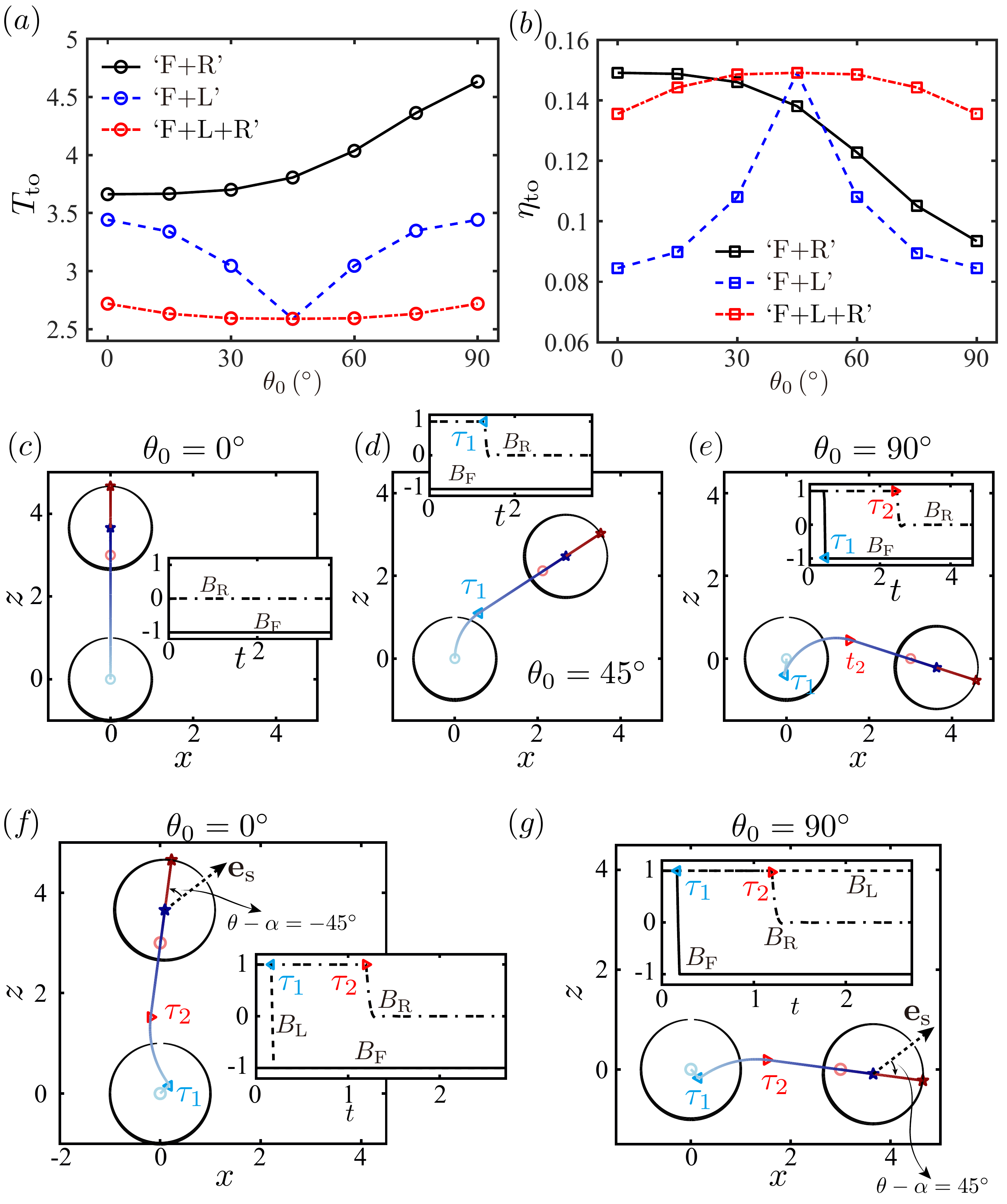}
\caption{Effects of the rotational mode $\BR$ of the squirmer on its time--optimal predatory performance. The (a) capture time $\Tot$, (b) predating efficiency $\etaot$ 
are plotted versus the initial orientation $\theta_0$ of the prey with respect to the predator. Three combinations of squirming modes, `F+R', `F+L' and `F+L+R' are adopted. Chasing dynamics of the `F+R' mode are illustrated for (c) $\theta_0=0^{\circ}$, (d) $\theta_0=45^{\circ}$ and (e) $\theta_0=90^{\circ}$; those of the `F+L+R' case are shown for (f) $\theta_0=0^{\circ}$ and (g) $\theta_0=90^{\circ}$. The insets plot the time evolution of the  modes. 
}
\label{fig:BR}
\end{figure}

In the above scenarios, the predator with a zero rotational mode $\BR=0$ does not rotate, hence its orientation remains in the $\be_z$ direction. To capture a prey at an arbitrary orientation, the predator must activate both the forward and lateral translational modes. However, if allowed to freely rotate by adopting a non--zero $\BR$ mode, the predator only needs one translational mode. We then ask how  such a combined rotational and translational mode compares with the combination of pure translational modes in the performance of TO predation. 
We show in Fig.~\ref{fig:BR} (a) the minimal capture time $\Tot$ and (b) the corresponding predatory efficiency $\etaot$ of the predator using three combinations of squirming modes, 1) forward plus rotational, `F+R', 2) `F+L', and 3) forward plus lateral plus rotational, `F+L+R'. The minimal capture time $\Tot$ of these three combinations diminishes in order regardless of the prey's orientation $\theta_0$ with respect to the predator. Besides, the `F+L+R' combination outperforms the other two in efficiency for most range of $\theta_0$. Moreover, for the `F+R' combination, $\Tot$ (resp. $\etaot$) increases (resp. decreases) with $\theta_0$ 
monotonically. This trend is in stark contrast with the symmetric (about $\theta_0=45^{\circ}$) profiles of $\Tot(\theta_0)$ and $\etaot(\theta_0)$ for the `F+L' and `F+L+R' combinations. 

We then analyse the detailed chasing dynamics for a better understanding. We first illustrate in Fig.~\ref{fig:BR} (c--e) how the `F+R' predator chases its prey oriented at $\theta_0 = 0^{\circ}$, $45^{\circ}$ and $90^{\circ}$. Intuitively, to reach an arbitrarily oriented prey, the predator using one other than two translational modes has to rotate to align its swimming direction exactly towards the prey.
For a special case when the prey with $\theta_0=0^{\circ}$ is initially ahead of the predator, no rotational motion is required for the predator as shown in Fig.~\ref{fig:BR}(c). Besides, when $\theta_0=45^{\circ}$, the predator adopts a rotational mode $\BR=1$ in full gear during $t\in[0,\tau_1]$ and then switches it off to swim straight towards the prey during $t>\tau_1$. 
It is revealed that for a non--zero $\theta_0$, the capture time comprises two parts: the first for rotational orientation and the second for straight swimming. The first part for orientation clearly increases with the initial angular difference $\theta_0$, which explains the monotonically increasing capture time $\Tot$ with $\theta_0$.
A less intuitive scenario occurs for $\theta_0=90^{\circ}$ when the prey is initially on the predator's right side: first, the predator moves backward and rotates rightward simultaneously during $t<\tau_1$, with both modes in full gear; then it stops the backward translation, while maintaining the full right rudder till $t=\tau_2$ when it exactly faces the prey; finally, the predator swims straightforward to the prey. The initial backward movement of predator seems awkward, which tends to retard predating by lengthening the predator--prey distance at the first glance. In fact, this seemingly awkward strategy embodies 
the wisdom of retreating in order to advance.
Moving backward actually reduces the angle $\theta-\alpha$ (that is  $\theta_0$ at $t=0$ shown in  Fig.~\ref{fig:sketch}(b)) between the predator--prey displacement and the predator's orientation, thus decreasing the time needed for orientation to effectively achieve a net time saving.

As discussed above,
replacing the lateral mode by the rotational mode, namely, shifting from the `F+L' to `F+R' combination results in asymmetric distributions of $\Tot(\theta_0)$ and $\etaot(\theta_0)$ about $\theta_0=45^{\circ}$. As reflected by the distinctive chasing dynamics for $\theta_0=0^{\circ}$, $\theta_0=45^{\circ}$ and $90^{\circ}$, this asymmetry is caused by the predator's necessity for a rotational orientation to face the prey. Hence, we postulate that the `F+L+R' squirming predator might exhibit similar asymmetric profiles of $\Tot(\theta_0)$ and $\etaot(\theta_0)$ owing to the rotational mode at play. In reality, this postulation is disproved by the symmetric profiles shown in Fig.~\ref{fig:BR}(a) and (b), which can be elucidated by scrutinizing how the `F+L+R' predator chases its prey initially at $\theta_0=0^{\circ}$ and $90^{\circ}$ as depicted by Fig.~\ref{fig:BR}(f) and (g), respectively. The set of trajectories of the predator and prey for $\theta_0=0^{\circ}$  matches that for $\theta_0=90^{\circ}$ in shape, and applying a $90$--degree rotational transformation allows them to overlap each other. In contrast to the `F+R' predator rotating to exactly face the prey when $\theta-\alpha=0^{\circ}$, this `F+L+R' predator also orients itself but instead to face the prey sitting in its northwest ($\theta-\alpha=-45^{\circ}$) or northeast ($\theta-\alpha=45^{\circ}$) direction before swimming straight towards the prey. This particular magnitude of $45^{\circ}$ 
enables the predator to exploit its maximum translational speed, $\sqrt{(\BFmax)^2+(\BLmax)^2}=\sqrt{2}$, to reach the prey in the post--rotation period $t>\tau_2$. This maximum translational speed exploited by the `F+L+R' predator leads to its faster predation compared to the `F+R' and `F+L' counterparts as shown in Fig.~\ref{fig:BR}(a). Indeed, the latter two can translate at a maximum velocity of $1$ other than $\sqrt{2}$. Besides, we comment that the difference between the initial bearing $\theta_0=0^{\circ}$ and $90^{\circ}$ explains the clockwise and counter--clockwise rotations of the predator, resulting in $\theta-\alpha=-45^{\circ}$ and $45^{\circ}$ respectively. This reasoning also justifies the $90$-degree rotational mapping between the two sets of trajectories associated with the two bearings.

\section{Results: reinforcement learning for a 
point prey and a finite--size prey}\label{sec:results_RL}
\subsection{RL--based optimization in the case of a point prey}
\begin{figure}
\centering
\includegraphics[scale=0.45]{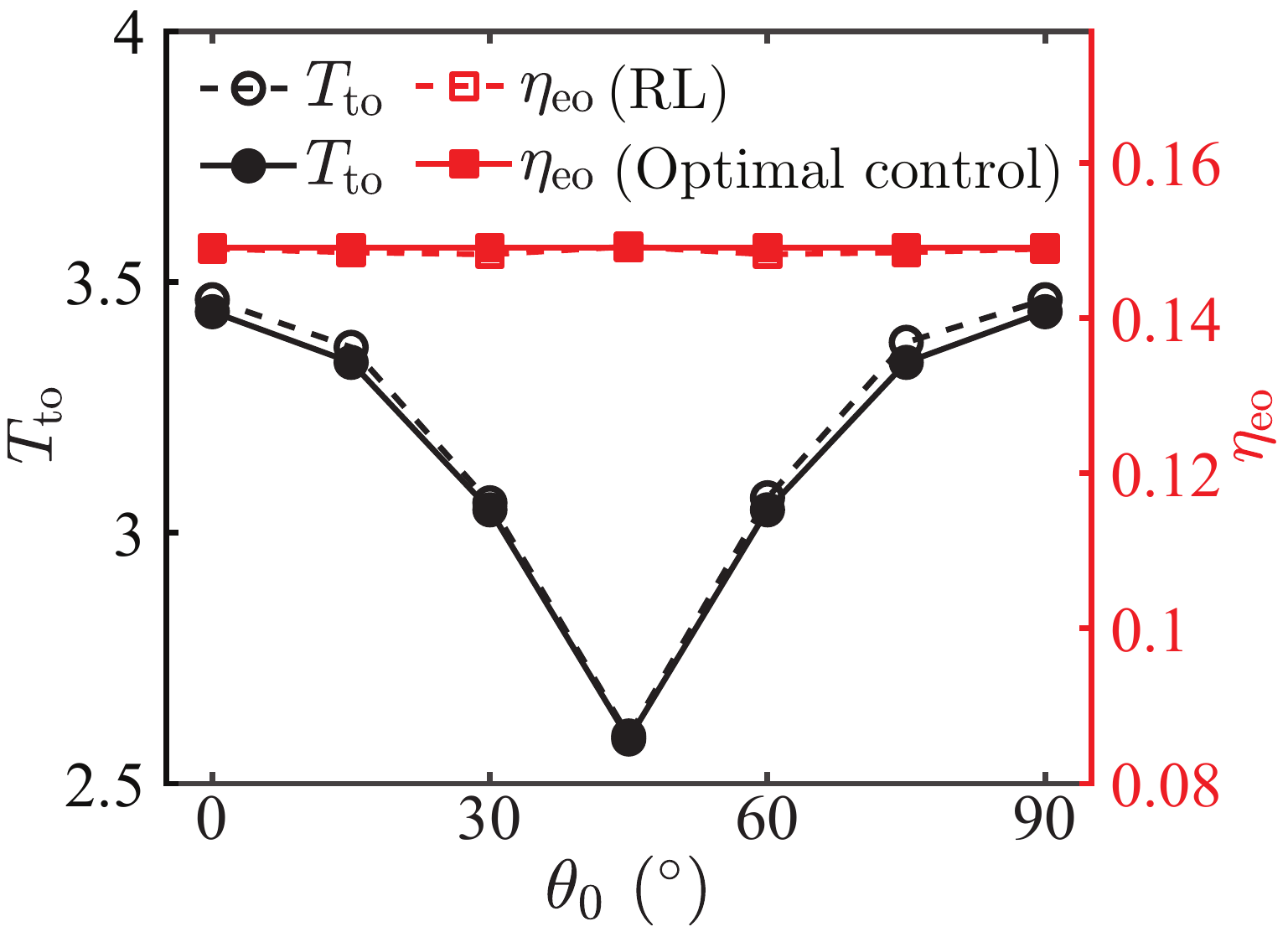}
\caption{Using RL to obtain the minimal capture time $\Tot$ and maximum efficiency $\etaoe$ of an `F+L' predator chasing a point prey. The results are compared against those based on the optimal control approach.
}
\label{fig:RL_OT_time}
\end{figure}

\begin{figure}
\centering
\includegraphics[scale=0.45]{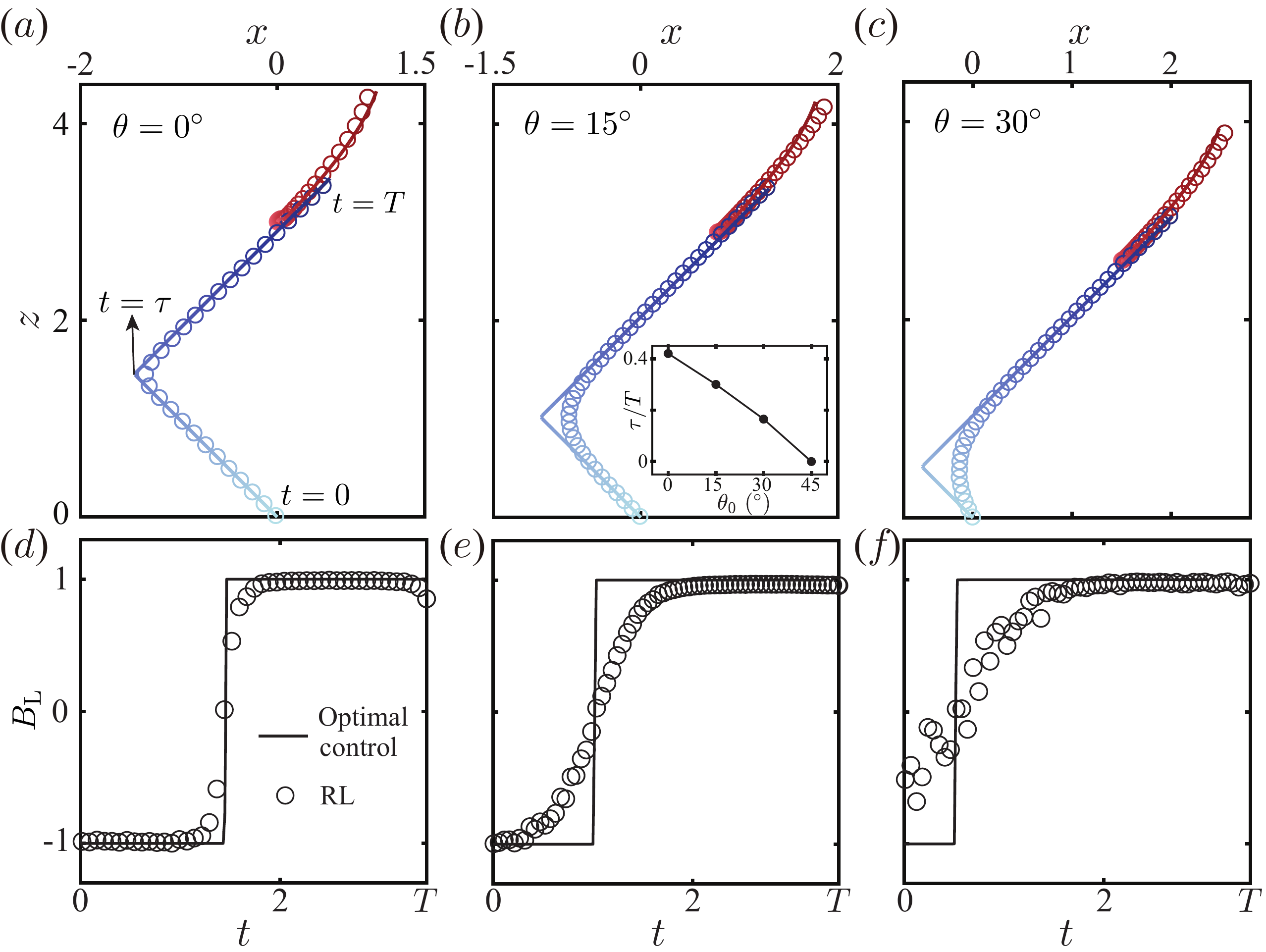}
\caption{Trajectories of the TO predators (blue) and their point prey (red) with an initial bearing of (a) $\theta=0^{\circ}$, (b) $15^{\circ}$ and (c) $30^{\circ}$, based on optimal control (curves) and RL (symbols). The time evolution of the lateral mode (action) $\BL$ of the predators are shown in (d), (e) and (f), respectively. The forward mode $\BF \approxeq -1$ as in Fig.~\ref{fig:OT_OE}(a) for both methods, and is thus not shown here.
The inset of (b) shows $\tau/T$ versus $\theta_0$ of the optimal control solutions.
} 
\label{fig:RL_OT_traj}
\end{figure}
To substantiate our study, we adopt an RL scheme to seek the optimal predatory strategies for a predator limited to the `F+L' squirming modes. Before studying a finite--size prey ($\chi>0$), we first address the scenario of a point prey ($\chi=0$), where the optimal solutions based on the optimal control approach (see Fig.~\ref{fig:OT_OE}) can be regarded as the globally--optimal solutions for benchmarking. As shown 
in Fig.~\ref{fig:RL_OT_time}, both the minimal capture time $\Tot$ and maximum efficiency $\etaoe$ obtained by RL agree well with their counterparts by optimal control. As also expected, a close inspection of Fig.~\ref{fig:RL_OT_time} shows that 
RL performs slightly worse than the optimal control, which further implies that the latter has indeed provided the globally--optimal solutions. Further comparing in Fig~\ref{fig:RL_OT_traj} the learned trajectories of the TO predator and its prey to their counterparts based on optimal control, we observe the RL--trained OT predator learns to execute a two--fold zigzag path identified by the optimal control approach. Especially, when $\theta_0=0^{\circ}$, the trajectories obtained from the two approaches almost collapse on each other; the sharp turn in the trajectory and the associated steep jump in the lateral model (action) $\BL$ are quantitatively captured by RL, as shown in  Fig.~\ref{fig:RL_OT_traj}(a) and (d).
However, as the initial bearing $\theta_0$ increases to $15^{\circ}$ and $30^{\circ}$, the RL solutions can only qualitatively reproduce the two--fold path, but fail to capture its sharp turn or the sudden jump of the swimming action  (see Fig.~\ref{fig:RL_OT_traj}(b), (c), (e) and (f)).
In fact, the degrading performance of RL at a larger bearing angle $\theta_0 < 45^{\circ} $ can be rationalized. For a two--fold path identified by optimal control, we define its two--foldedness $\tau/T$ as the time $\tau$ when the predator sharply turns scaled by the capture time. The foldedness decreases monotonically with $\theta_0\in \left[0, 45\right]^{\circ}$, becoming zero at $\theta_0=45^{\circ}$ corresponding to a straight chasing path (see the inset of Fig.~\ref{fig:RL_OT_time}(b)). This trend implies that time saving gained by executing a two--fold path diminishes with increasing $\theta_0$; or in another word, the extra time required by executing the straight path as a sub--optimal solution instead of the globally--optimal version decreases with growing $\theta_0<45^{\circ}$. Hence, at a sufficiently large $\theta_0$ featuring a negligible difference in the capture time between the sub-- and globally--optimal strategies, it becomes challenging for the RL algorithm to pinpoint exactly the globally--optimal one. 

Now back to Fig.~\ref{fig:RL_OT_traj}, strictly speaking, we shall not regard our RL--trained strategies as globally--optimal when $\theta_0=15^{\circ}$ and $30^{\circ}$. On the other hand, they indeed capture the essential features---two--fold path of the globally--optimal solutions. This promising observation thus motivates us to employ RL to optimize the predating strategy of a squirmer chasing a finite--size prey when the globally--optimal solutions are not available, as we will present in Sec.~\ref{sec:finite-size}. We add three more comments before proceeding forward. First, despite that the RL solutions deviate from the globally--optimal ones at increasing $\theta_0<45^{\circ}$, when $\theta_0=45^{\circ}$ featuring a globally--optimal straight path depicted in Fig.~\ref{fig:OT_OE}(e), RL can again exactly reproduce this solution. Also as in perfect agreement with the optimal control approach, RL can identify the straight paths reaching the optimal efficiency regardless of the initial bearing of the prey. These optimal straight paths are not shown here. Second, we have realized the crucial role of $\Gamma$ (in Eq.\eqref{eq:R_r_Gamma}) that represents a positive reward upon the capture time. Without this reward, the RL approach results in straight chasing trajectories regardless of the initial orientation $\theta_0$ of the prey, evidently being trapped in locally optima. Third, despite the substantial amount of work applying RL to optimize the locomotory gaits or path planning of different swimmers, to the best of our knowledge, we are not aware of an individual work that suggests that the RL--trained swimming strategy represents or resembles the globally--optimal one. It is indeed well--known that RL is easily trapped to local optima~\citep{sutton2018reinforcement,liepins1991deceptiveness,lehman2011novelty}.
The only exception might be the very recent work~\citep{nasiri2022reinforcement} having used RL to find asymptotically optimal navigating strategies of a point swimmer, which closely replicate the globally--optimal solutions identified previously by the same corresponding author with his collaborators~\citep{daddi2021hydrodynamics}. 
Together with \citet{daddi2021hydrodynamics,nasiri2022reinforcement}, 
our work indicates that RL-based optimization of swimming gaits or paths can be trapped to locally--optimal solutions, but it can also identify the global optima by adjusting the scheme properly. 

\subsection{RL--based optimization in the case of a finite--size prey ($\chi >0$)}\label{sec:finite-size}
\begin{figure}
\centering
\includegraphics[scale=0.37]{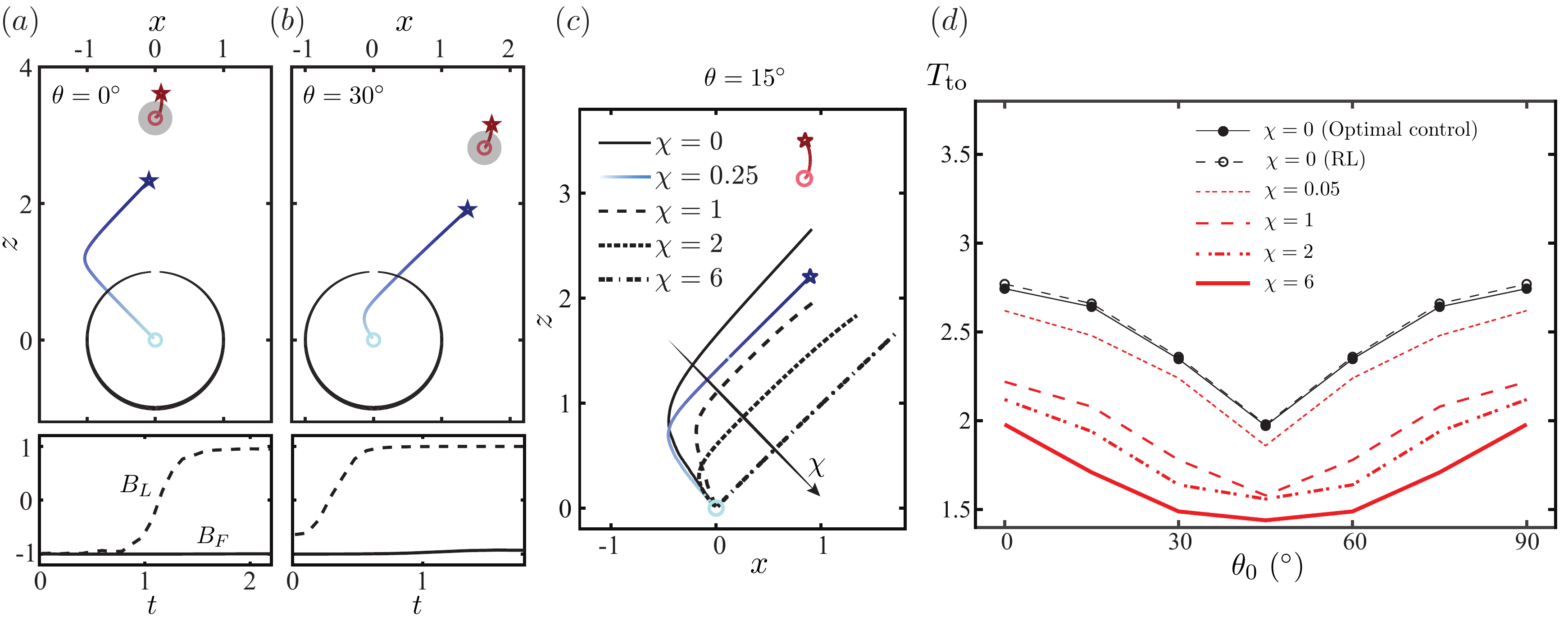}
\caption{Trajectories of the RL--trained TO `F+L' predator and its finite--size spherical prey of   radius $\chi=0.25$  initially located at (a) $\theta=0^{\circ}$ and (b) $30^{\circ}$; the lateral $\BL$ and forward $\BF$ modes of the squirmer are depicted in the bottom panels. The circles (resp. stars) denote the initial (resp. final) positions of the predator and prey.
(c) shows the dependence of the TO trajectory on the prey size $\chi\in[0, 6]$ when $\theta_0=15^{\circ}$.  (d), minimal capture time $\Tot$ versus the initial bearing $\theta_0$ and size $\chi$ of the prey. $\chi=0$ corresponds to a point prey studied via both optimal control and RL, whereas the results for a  finite--size prey ($\chi>0$) are obtained by RL. For all cases here, the cut--off distance is $\varepsilon = 5\times 10^{-2}$ and the initial surface--to--surface distance between the predator and prey is $|d_0| = 2$ as in Sec.~\ref{sec:results}.}
\label{fig:finite-chi}
\end{figure}
Here, we use RL to optimize the predatory strategies of an `F+L' squirmer
for capturing a finite-size spherical prey of dimensionless radius $\chi > 0$.
For the BIM--based RL environment, we use a cut--off  distance $\epsilon=5\times10^{-2}$ larger the previous value $4\times 10^{-3}$ due to the degrading accuracy  of BIM as involving two sufficiently close  surfaces. The finite--size effect of prey  does not change the typical two--foldedness of the TO chasing path, as exemplified by a $\chi=0.25$ prey initially at $\theta=0^{\circ}$ and $30^{\circ}$
(see Fig.~\ref{fig:finite-chi} (a) and (b), respectively).  
By further examining in Fig.~\ref{fig:finite-chi} (c) the
$\chi$--dependent TO predatory trajectory for $\theta_0=15^{\circ}$, we observe its decreasing two--foldedness in the case of a larger prey. For a sufficiently large prey of $\chi=6$, the TO path eventually becomes straight. We here provide a  phenomenological understanding of this change. As discussed in Sec.~\ref{sec:point_FL}, the two--fold path executed by the TO `F+L' predator enables exploiting its propulsion--induced disturbance flow to adjust the advection of the prey. The effect, however, diminishes for a larger prey being more difficult to advect due to its increased hydrodynamic resistance coefficient. This trend shall be responsible for the decreasing two--foldedness of the TO predatory path with increasing $\chi$. Furthermore, we depict in Fig.~\ref{fig:finite-chi} (d) the dependence of the minimal capture time $\Tot$ on the prey size $\chi$. Despite the unchanged symmetric profile of $\Tot(\theta_0)$ regardless of the varying $\chi$, the capture time $\Tot$ declines monotonically with the prey size. This negative relation can be rationalized by realizing the disturbance flow of the `F+L' squirmer, in overall, expels the prey away from it, as evidenced by the prey's path [see Fig.~\ref{fig:RL_OT_traj} and Fig.~\ref{fig:finite-chi}(a) and (b)]. Therefore, a larger expelled prey  travels a shorter distance  reducing the capture time. We also note that the wide range of chosen $\chi$ has been motivated by the various realistic scenarios introduced in Sec.~\ref{sec:intro}.

\section{Conclusion}
In this work, we study, in the creeping flow regime, a swimming predator modelled by a spherical squirmer chasing a non--motile point or finite--size spherical prey advected by the disturbance flow generated by the former. Using optimal control for a point prey and RL for general situations, we optimize the predatory strategies of the squirmer that can translates forward (`F') and laterally (`L'), generates a stresslet (`S') flow or rotates (`R') in the fluid.
We have identified the best time sequences of the squirming modes to achieve time--optimal (TO) or efficiency--optimal (EO) predating, namely to minimize the capture time or to maximize the predatory efficiency, respectively.   

We first focus on a point prey. The EO `F+L' predator swims straight towards the prey regardless its initial bearing with respect to the predator. In contrast, the TO predator follows an L--shaped route, hence travelling a longer distance than the EO predator. This chasing strategy can be understood by examining how the disturbance flow of the predator advects the prey: the EO predator generates a flow that persistently expels the prey away from it; the TO counterpart has been optimized to adapt its position with respect to the prey, such that its disturbance flow can be harnessed to advect the latter towards itself to some extent. This peculiar route may not be easily revealed intuitively. Besides, we show that incorporating an additional stresslet mode of magnitude $\BSmax=1$ allows the `F+L+S' EO predator to considerably outperform the `F+L' counterpart in every aspects of the predatory performance; in most cases, the former captures the prey faster, consumes less energy, travels a shorter distance, and gains a higher predatory efficiency (see Fig.~\ref{fig:BS}). We recall that for an isolated  squirmer, introducing the stresslet mode does not change its speed, however increasing the energy expenditure and decreasing its efficiency~\citep{blake1971spherical}. For predation, the counterintuitive energy--saving and efficiency enhancement result from the predator's largely reduced capture time and travelling distance. This reduction is mostly pronounced when the prey is initially ahead of the predator, which achieves so by utilizing the stresslet flow to suck the prey towards itself. A similar scenario was revealed by~\citet{tam2011optimal} for a biflagellated swimmer that exploits its  strokes--induced  currents to achieve optimal nutrients uptake. Then, we have examined how the maximum magnitude $\BSmax$ of the stresslet mode influences the `F+L+S' squirmer optimized for the fastest predation. Increasing the magnitude reduces the predator's capture time and travelling distance as expected,  however significantly increases the consumed energy. The two competing trends result in the non--monotonic variation of the predatory efficiency versus $\BSmax$; accordingly, the TO predator attains the highest predatory efficiency at an optimal value $\BSmax \approx 0.7-0.8$. This finding might help formulate useful guidelines for designing micro--robots. In addition,
we have also investigated  the potential role of rotational motion in the TO predation. Compared to a translating `F+L' predator, the `F+R' counterpart combining the forward translation and rotation spends more time catching the prey; while the `F+L+R' squirmer using two translational motions and rotation seizes the prey faster than the former two compeers. Unlike the `F+L' TO predator following an L--shaped path, the `F+R' compeer first rotates to face the prey exactly and then swims straight to it (see Fig.~\ref{fig:BR}(d)). Thus, the total capture time comprises two parts---one used for rotational reorientation and the other for straight chasing. For a prey initially exactly on the right side requiring a considerable rotation, the TO predator has been optimized to adopt a non--intuitive strategy of retreating in order to advance: it first swims backward, leaving other than approaching the prey in appearance; this trick effectively reduces the angular difference between its orientation and the prey's bearing and thus the corresponding time for rotation, leading to a net time saving.

Besides optimal control, we have used RL to seek the optimal  strategies of an `F+L' squirmer chasing a point prey or a spherical one of radius $\chi>0$. For the latter, a BIM solver is developed to emulate the RL environment. For a point prey, we have demonstrated that our RL--based solutions qualitatively (even quantitatively in some cases) agree with the globally--optimal ones from optimal control. In particular, the former capture the  nonintuitive two--foldedness of the TO path of the latter. Applying RL to a spherical prey of radius $\chi>0$, we have identified that the two--foldedness of the TO path decreases with increasing $\chi$, and the TO path eventually becomes straight at a sufficiently large $\chi$. We have also shown that the minimal capture time decreases monotonically with $\chi$ because a larger prey is more difficult to be advected by the predator.

\end{document}